\documentclass[onecolumn]{emulateapj}
\usepackage{amsmath}     
\usepackage{wasysym}           
\usepackage{graphicx}
\usepackage{amssymb}
\usepackage{epstopdf}
\usepackage{mathrsfs}
\usepackage[T1]{fontenc}
\usepackage{natbib}

\begin{document}
\title{The general relativistic equations of radiation hydrodynamics in the viscous limit}
\author{Eric R. Coughlin\altaffilmark{1} and Mitchell C. Begelman\altaffilmark{1}}
\affil{JILA, University of Colorado and National Institute of Standards and Technology, UCB 440, Boulder, CO 80309}
\email{eric.coughlin@colorado.edu, mitch@jila.colorado.edu}
\altaffiltext{1}{Department of Astrophysical and Planetary Sciences, University of Colorado, UCB 391, Boulder, CO 80309}

\begin{abstract}
We present an analysis of the general relativistic Boltzmann equation for radiation, appropriate to the case where particles and photons interact through Thomson scattering, and derive the radiation energy-momentum tensor in the diffusion limit, with viscous terms included. Contrary to relativistic generalizations of the viscous stress tensor that appear in the literature, we find that the stress tensor should contain a correction to the comoving energy density proportional to the divergence of the four-velocity, as well as a finite bulk viscosity. These modifications are consistent with the framework of radiation hydrodynamics in the limit of large optical depth, and do not depend on thermodynamic arguments such as the assignment of a temperature to the zeroth-order photon distribution. We perform a perturbation analysis on our equations and demonstrate that, as long as the wave numbers do not probe scales smaller than the mean free path of the radiation, the viscosity contributes only decaying, i.e., stable, corrections to the dispersion relations. The astrophysical applications of our equations, including jets launched from super-Eddington tidal disruption events and those from collapsars, are discussed and will be considered further in future papers.
\end{abstract}

\keywords{radiation: dynamics -- radiative transfer -- relativistic processes}

\section{Introduction}
Radiation contributes substantially to the dynamics of many astrophysical systems. The most natural way to analyze the mechanics of such systems is through the formalism of radiation hydrodynamics, wherein one treats the radiation as a fluid that interacts with matter. With the covariant derivative denoted $\nabla_{\mu}$, the equations of radiation hydrodynamics can be generalized to include both gravitational fields and relativistic motions by writing them in the manifestly covariant form

\begin{equation}
\nabla_{\mu}(T^{\mu\nu}+R^{\mu\nu}) = 0 \label{radhydro},
\end{equation}
where Greek indices range from $0$ to 3, repeated upper and lower indices imply summation,

\begin{equation}
T^{\mu\nu} = w'U^{\mu}U^{\nu}+p'g^{\mu\nu}
\end{equation}
is the energy-momentum tensor of the massive constituents of the fluid ($w'$ is the enthalpy in the fluid rest-frame, $U^{\mu}$ is the four-velocity, $p'$ is the pressure in the fluid rest-frame, and $g^{\mu\nu}$ is the inverse of the metric associated with the background geometry) and

\begin{equation}
R^{\mu\nu} = \int{}k^{\mu}k^{\nu}f\frac{d^3k}{k^{0}} \label{radmom}
\end{equation}
is the energy-momentum tensor of the radiation (see, e.g., \citealt{mih84} for a more thorough discussion of the origin of this tensor). In equation \eqref{radmom}, $k^{\mu}$ is the four-momentum of a photon, $f$ is the distribution function of the radiation that describes the density of quanta in both momentum and position space, and $d^3k/k^{0}$ is the Lorentz-invariant phase-space volume. Because $f$ is a scalar, such a ``moment" formalism, i.e., proceeding by taking integrals over the distribution function, is a natural way to analyze the dynamics of the radiation in a covariant fashion. Note, however, that equation \eqref{radmom} is only valid in a locally-flat frame -- one in which $g_{\mu\nu} = \eta_{\mu\nu}$, $\eta_{\mu\nu}$ being the Minkowski metric -- as otherwise one must include factors that depend on the metric (see, e.g., \citealt{deb09a}). To ensure that the radiation energy-momentum tensor (and derivatives thereof) transforms correctly at all points in the spacetime under consideration, one can explicitly insert the metric-dependent factors that enter into the phase-space volume element. Another, equally valid manner by which one can obtain the general-relativistic form of $R^{\mu\nu}$, however, is by evaluating the integrals in equation \eqref{radmom} and writing the results in a manifestly-covariant form, guaranteeing their frame independence. Because we will be working directly with the distribution function, the latter route is the simpler one to follow and is the one that we will pursue in our ensuing analyses (see sections 3 and 4).

For many astrophysical applications of the equations of radiation hydrodynamics, the medium under consideration is optically thick, meaning that the radiative flux observed at any location within the fluid is very nearly zero. This means, equivalently, that a photon is scattered a large number of times as it propagates through the medium, and that the radiation field as seen by a moving fluid element within the medium is approximately isotropic. It has been known for some time, however, that the finite mean free path of a photon leads to the presence of viscous-like terms in the radiation energy-momentum tensor \citep{tho30}. The viscous nature of some fluids, especially in applications for which photons dominate the pressure of a system, can therefore be attributed in part to their interactions with radiation.

\citet{eck40} analyzed the origin of generic, viscous effects in relativistic fluids, whether due to radiation or to other phenomena (see also \citealt{lan58}). He exploited the fact that the four-velocity of the fluid, $U^{\mu}$, and the projection tensor, $\Pi^{\mu\nu} = U^{\mu}U^{\nu}+g^{\mu\nu}$, are time-like and space-like tensors, respectively, that can be used to decompose any arbitrary tensor. Eckart realized that each of the terms in his decomposition had a physical interpretation, related to the dynamic viscosity or heat conduction of the fluid, which allowed him to postulate a form for the viscous stress tensor of a relativistic fluid that reduced correctly to its non-relativistic counterpart. 

\citet{tho30} (and others after him, e.g., \citealt{lin66, cas72, buc79, mun86, che00}) used the special relativistic Boltzmann equation, which describes changes to the distribution function due to particle-particle collisions, to derive the correction to the radiation energy-momentum tensor for the special case of Thomson scattering. Eckart's approach, on the other hand, was more phenomenological in nature, employing thermodynamic arguments and an understanding of the Newtonian limit of the viscous stress tensor to derive its relativistic generalization. \citet{wei71}, with the intent of using the results to analyze entropy production in the early universe, compared the two approaches and showed that one of Eckart's assumptions, namely that the viscous stress tensor be trace free, led to the incorrect conclusion that the bulk viscosity of the fluid vanish (see also \citet{mis65}, where a similar approach is used to evaluate radiative effects during core-collapse supernovae). In his analysis, the part of the relativistic stress tensor expressing viscosity and heat conduction, a slightly generalized version of Eckart's, was of the form

 \begin{equation}
\Delta{T^{\mu\nu}} = -\eta\,\Pi^{\mu\sigma}\Pi^{\nu\rho}\bigg{(}\nabla_{\sigma}U_{\rho}+\nabla_{\rho}U_{\sigma}-\frac{2}{3}g_{\sigma\rho}\nabla_{\alpha}U^{\alpha}\bigg{)}-\zeta\,\Pi^{\mu\nu}\nabla_{\alpha}U^{\alpha} 
-\chi\bigg{(}\Pi^{\nu\sigma}U^{\mu}+\Pi^{\mu\sigma}U^{\nu}\bigg{)}\bigg{(}TU^{\rho}\nabla_{\rho}U_{\sigma}+\nabla_{\sigma}T\bigg{)} \label{deltaT},
\end{equation}
where $\eta$, $\zeta$, and $\chi$ are the coefficients of dynamic viscosity, bulk viscosity, and heat conduction, respectively, and $T$ is the temperature of the gas. In the special-relativistic limit, $\nabla_{\mu} = \partial_{\mu}$ is just the partial derivative, but is the covariant derivative if curvilinear coordinates are being used (or if the fluid is in a gravitational field). His comparison between the two theories enabled him to calculate $\eta$, $\zeta$, and $\chi$ for a radiating fluid, confirming the notion that small-scale anisotropies in the radiation field, i.e., on the order of the mean free path of a photon, generate viscous-like effects. 

Although they yield similar results, the fluid/thermodynamic approach of \citet{eck40}, and correspondingly that of \citet{wei71}, is fundamentally different from the kinetic theory approach of \citet{tho30}. The first difference arises from the fact that the former is a single-fluid analysis, meaning that $\Delta{T^{\mu\nu}}$, as given by equation \eqref{deltaT}, is the viscous correction to ``the fluid." The second disparity comes about because Eckart defines thermodynamic quantities in terms of the energy-momentum tensor of the fluid; for example, \citet{eck40} and \citet{wei71} both \emph{define} the comoving energy density as $e' = U_{\mu}U_{\nu}T^{\mu\nu}$, meaning that there is no correction to the observed energy. It is for this reason that there is no term proportional to $U^{\mu}U^{\nu}$ in equation \eqref{deltaT}; however, the Eckart decomposition for an arbitrary tensor, one that has no restriction imposed upon it concerning the comoving energy density, will have such a term.

On the other hand, the kinetic theory approach considers the radiation and the matter to be two separate, interacting fluids -- the same viewpoint that underlies all of radiation hydrodynamics -- meaning that the viscous terms are understood as corrections to the radiation energy-momentum tensor, rather than that of the matter. Also, the kinetic theory description uses the Boltzmann equation to determine the distribution function, the viscous stress tensor then being written in terms of integrals over that function, viz., equation \eqref{radmom}. There is thus no need to define bulk physical parameters, such as the temperature $T$ or the comoving energy density $e'$, in terms of the stress tensor. Indeed, one could verify the validity of Eckart's assumption that $\Delta{e'} = U_{\mu}U_{\nu}\Delta{T^{\mu\nu}} = 0$ if one knew the distribution function. 

Our goal here is to analyze the Boltzmann equation and thereby evaluate the stress tensor for a relativistic, radiating fluid in the limit that the finite mean free path of the radiation provides the source of the viscosity. In section 2 we present the general relativistic Boltzmann equation, the formalism of general relativity being necessary because of the fact that scattering is handled most easily in the comoving (accelerating) frame of the fluid. In section 3 we restrict our attention to the case where the scattering is dominated by Thomson scattering and we solve the resultant equation for the distribution function to first order in the mean free path. Section 4 presents the equations of radiation hydrodynamics in the viscous limit and we demonstrate that the stress tensor departs from equation \eqref{deltaT} in a few important ways, the first being that our equation has an additional correction to the comoving energy density, and the second being that our result is independent of thermodynamic considerations, such as the assignment of a temperature, and is therefore applicable to non-equilibrium radiation fields. Our coefficient of bulk viscosity also differs from that derived by \citet{wei71}. We perform a Fourier analysis of the perturbed equations in section 5 and show that they are indeed stable to perturbations of the fluid on scales larger than the mean free path of the radiation. A discussion of future applications, some comments on current radiation magnetohydrodynamic codes and conclusions are presented in section 6.

\section{Relativistic Boltzmann equation}
The distribution function for the radiation must satisfy a transfer equation -- some statement of the conservation of photon number. As mentioned in the introduction, a natural choice for this equation is the Boltzmann equation, which describes the changes to the distribution function owing to emission, absorption, and scattering processes. However, because photons are massless, we must use a relativistic version of the Boltzmann equation; an obvious generalization to the relativistic regime is

\begin{equation*}
\frac{\partial{f}}{\partial{t}}+v^{i}\frac{\partial{f}}{\partial{x^{i}}}+\dot{v}^{i}\frac{\partial{f}}{\partial{v^{i}}} = \delta{f}\big{|}_{coll}
\end{equation*}
\begin{equation}
 \rightarrow 
k^{\mu}\frac{\partial{f}}{\partial{x^{\mu}}}+\dot{k}^{\mu}\frac{\partial{f}}{\partial{k^{\mu}}} = \delta{f}\big{|}_{coll}\label{1},
\end{equation}
where Latin indices adopt the range 1 -- 3, Greek indices range from 0 -- 3, repeated upper and lower indices imply summation, and $k^{\mu}$ is the four-momentum of a photon. The right-hand sides represent changes to the distribution function through interactions with the surrounding medium. Equation \eqref{1} is the relativistic Boltzmann equation often encountered in the literature (e.g., \citealt{mih84}). There are, however, two important subtleties associated with equation \eqref{1} that are often not mentioned explicitly. 

The first issue is that the sum over $\dot{k}^{\mu}$ on the left-hand side of \eqref{1} must incorporate the constraint that the four-momentum of the photon lie on the null cone, viz., $k_{\mu}k^{\mu} = 0$ (this problem vanishes in flat space because photons travel in straight lines, and hence $\dot{k}^{\mu} = 0$), which raises the question of whether or not the sum should occur over all four components of the momentum or just three. Furthermore, if we take the latter route, which three should we choose? Comparing equation \eqref{1} to the non-relativistic expression suggests taking the three spatial components; relativistic covariance, however, demands that the time component should not be treated differently from the spatial momenta. 

The second subtlety stems from the fact that, in general relativity, there are two different momenta from which we can choose to describe the system: the covariant, $k_{\mu}$, and the contravariant, $k^{\mu}$, components, related by $k_{\mu} = g_{\mu\nu}k^{\nu}$ where $g_{\mu\nu}$ is the metric. The distribution function treats the spatial variables, $x^{\mu}$, as independent of the momentum. Therefore, should we consider $x^{\mu}$ and $k^{\mu}$ as independent coordinates, or $x^{\mu}$ and $k_{\mu}$? It is apparent that, depending on which one we choose, the results will differ as the metric $g_{\mu\nu}$ is a function of the spatial coordinates. 

In our ensuing treatment of equation \eqref{1}, we will be analyzing interactions in the fluid frame -- the one comoving with a given fluid parcel. Such a frame will, in general, be accelerating, meaning that $\dot{k}^{\mu} \neq 0$ and there will be an acceleration-induced metric $g_{\mu\nu}$, forcing us to confront each of the previously-raised questions. Recently, \citet{deb09a, deb09b} addressed these issues directly by returning to the most general definition of the distribution function, a sum of Dirac delta functions in position and momentum space. They demonstrated \citep{deb09a} that one may consider either the covariant, spatial components $k_{i}$  or the contravariant components $k^{i}$ as the momentum independent from $x^{\mu}$. However, as we mentioned previously, the Boltzmann equations that result from these choices are not identical, meaning that one must also define two different distribution functions, one dependent on the covariant components and the other on the contravariant components, to proceed unambiguously. Even though the routes were shown to be equivalent, one must be careful to use the components appropriate to a given Boltzmann equation.

In \citet{deb09b}, the authors showed that equation \eqref{1} is correct if 1) the sum involving $\dot{k}^i$ in the second term is performed over the three spatial components, 2) the distribution function $f$ is considered to be a function of the contravariant, spatial components $k^{i}$, and 3) any appearance of $k_{0}$ (or $k^{0}$) is replaced by $k_{0}(k^{i})$. The dependence of $k_0$ on the spatial momenta can be determined by solving the equation $k_{\mu}k^{\mu} = 0$ for $k_0$. As we stated, the frame of interest is the comoving frame of the fluid, and we will denote the components of any tensor in this frame with a prime on the index; e.g., $k^{i'}$ is the $i$th component of the momentum in the comoving frame. With this convention, the relativistic Boltzmann equation becomes

\begin{equation}
k^{\mu'}\frac{\partial{f}}{\partial{x^{\mu'}}}-\Gamma^{i'}_{\mu'\nu'}k^{\mu'}k^{\nu'}\frac{\partial{f}}{\partial{k^{i'}}} = \delta{f}\big{|}_{coll} \label{boltzmanncomoving},
\end{equation}
where we have used the geodesic equation,

\begin{equation}
\dot{k}^{i'}+\Gamma^{i'}_{\mu'\nu'}k^{\mu'}k^{\nu'} = 0,
\end{equation}
to replace $\dot{k}^{i'}$, and

\begin{equation}
\Gamma^{i'}_{\mu'\nu'} = \frac{1}{2}g^{i'\alpha'}\bigg{(}\partial_{\mu'}g_{\nu'\alpha'}+\partial_{\nu'}g_{\mu'\alpha'}-\partial_{\alpha'}g_{\mu'\nu'}\bigg{)}
\end{equation}
are the Christoffel symbols associated with the metric $g_{\mu'\nu'}$ which, for our purposes, is induced by the acceleration associated with the comoving frame. 

Equation \eqref{boltzmanncomoving} was provided by \citet{lin66} and \citet{cas72}. Since they were considering spherically symmetric flows, they opted to change the form of equation \eqref{boltzmanncomoving} by using an orthonormal tetrad adapted to spherical coordinates. For our purposes, however, we will not be in the position to take advantage of any specific coordinate symmetries, so equation \eqref{boltzmanncomoving} will suffice. 

Finally, it should also be mentioned that the right-hand side is evaluated at a specific location -- the point in space and time where the collision occurs. The left-hand side, therefore, must also be evaluated at that spacetime point.

\section{Diffusion approach to the transport equation}
Here we will first make the approximation that the temperature of the gas is high enough such that all species, considered to comprise a single, massive fluid, are ionized, meaning that the right-hand side of equation \eqref{boltzmanncomoving} incorporates effects due only to scattering. In this case, the collisional term can be written \citep{hsi76}

\begin{equation}
\delta{f}\big{|}_{coll} = n'\int{}R(k^{\mu'},k^{\mu'}_i)f(x^{\mu'}, k^{\mu'}_i)\frac{d^3k'_i}{k^{0'}_i}-n'\int{}R(k^{\mu'}_f,k^{\mu'})f(x^{\mu'}, k^{\mu'})\frac{d^3k'_f}{k^{0'}_f} \label{scattering},
\end{equation}
where $R$ is the redistribution function (not to be confused with the stress-energy tensor) and $n'$ is the rest-frame number density of scatterers. The first term represents scatterings into the state $k^{\mu'}$ from any initial momentum state $k^{\mu'}_i$, while the second embodies scatterings out of state $k^{\mu'}$ into any other state $k^{\mu'}_f$. As long as photon wavelengths are long compared to the Compton wavelength and the gas is non-relativistic in the rest frame of the fluid, two suppositions we will make here, the redistribution function (in the comoving frame of the fluid) is that appropriate to Thomson scattering:

\begin{equation}
R(k^{\mu'},k^{\mu'}_i) = \frac{3\sigma_T}{16\pi}\bigg{(}1+(\hat{k}^{x'}\hat{k}_i^{x'}+\hat{k}^{y'}\hat{k}_i^{y'}+\hat{k}^{z'}\hat{k}_i^{z'})^2\bigg{)}\delta(k^{0'}-k^{0'}_i), \label{thomson}
\end{equation}
where $\sigma_T$ is the Thomson cross section (or, more generally, the cross section relevant to the scatterer). The equation for $R(k^{\mu'}_f, k^{\mu'})$ is identical to that for $R(k^{\mu'},k^{\mu'}_i)$ but with $k^{\mu'}_i \rightarrow k^{\mu'}_f$. Inserting equation \eqref{thomson} into equation \eqref{scattering} and the result of that substitution into equation \eqref{boltzmanncomoving} gives our final form for the transfer equation. 

It should be noted that the integrals in equation \eqref{scattering} are performed under the restriction that the photon four-momenta lie on the null cone. Since collisions occur instantaneously at fixed locations in space, we can approximate the metric to be locally that of flat space, i.e., $g_{\mu'\nu'} = \eta_{\mu'\nu'}$, and we can use the null cone condition ($k_{\mu'}k^{\mu'} = 0$) to write $k^{0'} = |\mathbf{k}'| = \sqrt{(k^{x'})^2+(k^{y'})^2+(k^{z'})^2}$. Every appearance of $k^{0'}$ in equation \eqref{scattering} can thus be replaced by $|\mathbf{k}'|$. 

Our goal here is to discern how the radiation field responds to gradients in the flow velocity. In the next section, we will use the equations of radiation hydrodynamics to deduce how the flow couples to changes in the radiation field, thereby completing the picture. To achieve this goal we will adopt a diffusion approximation, asserting that the distribution function may be written as $f \simeq f_0+f_1$, where $f_1$ is a small correction to $f_0$. The ``smallness" of $f_1$ is encoded in the mean free path $\sim (n'\sigma_T)^{-1}$ and the gradients of the flow velocity, $v$, across the mean free path, meaning that $f_1 \sim f_0\,d{v}/d\tau$, where $\tau \sim n'\sigma_T\,x$ is the optical depth.  

Because the right-hand side of equation \eqref{boltzmanncomoving} is proportional to the optical depth (per unit length), the consistency of our diffusion approach demands that $\delta{f_0}\big{|}_{coll} = 0$. It can be verified that the right-hand side vanishes for any function $f_0$ that depends only on the magnitude of the momentum, meaning that $f_0$ is isotropic in the comoving frame. This result is consistent with the expectation that, in an optically-thick medium, the flux of radiation observed in a frame comoving with a fluid parcel is very nearly zero. Since the collision operator acts at a point in spacetime, however, the zeroth-order distribution function can also contain any other secular variation in space and time, meaning that its most general form is $f_0(k^{\mu'},x^{\mu'}) = f_0(|\mathbf{k}'|,x^{\mu'})$.  The first-order transfer equation, to be solved for $f_1$, is then

\begin{equation}
k^{\mu'}\frac{\partial{f_0}}{\partial{x^{\mu'}}}-\Gamma^{i'}_{\mu'\nu'}k^{\mu'}k^{\nu'}\frac{\partial{f_0}}{\partial{k^{i'}}} = \delta{f_1}\big{|}_{coll}.
\end{equation}

To make more progress on this relation, we must determine the metric associated with the comoving frame. To do so, we will first assume a planar configuration of the fluid, with neither velocity nor variation in the $x$ direction. We will then use the fact that the coordinate transformation to move into the comoving frame of a given fluid parcel is a local Lorentz transformation, the inverse of which is given by

\begin{equation}
t = \int_0^{t'}\Gamma\,dt''+\int_0^{z'}\Gamma{v_z}\,dz''+\int_0^{y'}\Gamma{v_y}\,dy'' \label{ttransform},
\end{equation}
\begin{equation}
z = \int_0^{t'}\Gamma{v_z}\,dt''+\int_0^{z'}\bigg{(}1+\frac{v_z^2}{v^2}\bigg{(}\Gamma-1\bigg{)}\bigg{)}\,dz''+\int_0^{y'}\frac{v_zv_y}{v^2}\bigg{(}\Gamma-1\bigg{)}dy'',
\end{equation}
\begin{equation}
y = \int_0^{t'}\Gamma{v_y}\,dt''+\int_0^{z'}\frac{v_zv_y}{v^2}\bigg{(}\Gamma-1\bigg{)}\,dz''+\int_0^{y'}\bigg{(}1+\frac{v_y^2}{v^2}\bigg{(}\Gamma-1\bigg{)}\bigg{)}\,dy'',
\end{equation}
\begin{equation}
x = x' \label{xtransform},
\end{equation}
where $v^2 = v_z^2+v_y^2$, $\Gamma = (1-v^2)^{-1/2}$ is the Lorentz factor (not the Christoffel symbol), and we have, without loss of generality, chosen the origin of the primed coordinate system to coincide with that of the lab frame. The integrals are necessary here because the velocities are all dependent on the coordinates $t'$, $z'$, and $y'$, but because we will ultimately be evaluating our expressions at the origin, one would obtain the same answer by letting $\int_{0}^{t'}\Gamma\,dt'' = \Gamma\,t'$, etc. Double-primed coordinates are simply dummy variables where for each integrand we let $t' \rightarrow t''$, etc. The line element, which in flat space is given by

\begin{equation}
ds^2 = -dt^2+dz^2+dy^2+dx^2 \label{line},
\end{equation}
is invariant with respect to our choice of coordinates; the metric can therefore be determined by differentiating equations \eqref{ttransform} -- \eqref{xtransform}, inserting the results into equation \eqref{line} and grouping terms (see \citealt{cas72} for a similar, but non-relativistic, approach).  

Calculating the Christoffel symbols, using the chain rule to determine $\partial{f}/\partial{k^{i'}}$ and $\partial{f}/\partial{x^{\mu'}}$, inserting the expressions into equation \eqref{boltzmanncomoving} and evaluating the result at the origin (as this is the location of the fluid parcel -- where the scattering occurs and where spacetime is locally Minkowskian), we find

\begin{equation*}
k^{\mu'}\frac{\partial{f_0}}{\partial{x^{\mu'}}}-\Gamma^{i'}_{\mu'\sigma'}k^{\mu'}k^{\sigma'}\frac{\partial{f_0}}{\partial{k^{i'}}} = 
\end{equation*}
\begin{equation}
k^{\mu'}\frac{\partial{f_0}}{\partial{x^{\mu'}}}\bigg{|}_{|\mathbf{k'}|}-\frac{\Gamma^2}{v^2}\frac{\partial{f_0}}{\partial|\mathbf{k}'|}\bigg{(}A_1(k^{z'})^2+A_2(k^{y'})^2+A_3|\mathbf{k}'|k^{z'}+A_4|\mathbf{k}'|k^{y'}+A_5k^{y'}k^{z'}\bigg{)}, \label{boltzmannlhs}
\end{equation}
where, by using the definitions of the Christoffel symbols and using $U^{\mu'} = \delta^{\mu'}_{\,\,0'}$, it can be shown that 

\begin{equation}
\frac{\Gamma^2}{v^2}A_1 = \nabla_{z'}U_{z'} \label{A1},
\end{equation}
\begin{equation}
\frac{\Gamma^2}{v^2}A_2 = \nabla_{y'}U_{y'},
\end{equation}
\begin{equation}
\frac{\Gamma^2}{v^2}A_3 = \nabla_{0'}U_{z'},
\end{equation}
\begin{equation}
\frac{\Gamma^2}{v^2}A_4 = \nabla_{0'}U_{y'},
\end{equation}
\begin{equation}
\frac{\Gamma^2}{v^2}A_5 = \nabla_{y'}U_{z'}+\nabla_{z'}U_{y'} \label{A5}.
\end{equation}
We broke up the derivative $\partial/\partial{x^{\mu'}}$ into two separate components: $\partial{f_0}/\partial{|\mathbf{k'}|}$, taken such that all appearances of $x^{\mu'}$ not contained in the definition of $|\mathbf{k}'| = \sqrt{g_{i'j'}k^{i'}k^{j'}}$, through the metric, are kept constant, and $\partial{f_0}/\partial{x^{\mu'}}\big{|}_{|\mathbf{k}'|}$, taken such that all spatial coordinates that appear through $|\mathbf{k}'|$ are held fixed. Note that we must use the relativistically-correct version of the magnitude of the photon momentum, i.e., one involving the metric, because we are taking derivatives of the distribution function before evaluating the result at the location of the fluid parcel. Thus, even though spacetime is flat exactly at the point of interest, deviations exist at neighboring locations -- the derivative requiring that we evaluate the distribution function at those locations. Though we did not explicitly denote it, the derivatives in equations \eqref{A1} -- \eqref{A5} are to be evaluated at the origin.

The right-hand side of equation \eqref{boltzmanncomoving} involves an integral over $f_1$ (recall that $R(k^{\mu'},k^{\mu'}_i)$ is given by equation \eqref{thomson}), and the most direct means of evaluating $f_1$ would be to expand both the left-hand side of the transfer equation and the function $f_1$ in terms of spherical harmonics. Instead of pursuing this route, however, we will make the educated guess

\begin{equation}
f_1 = B_1(k^{z'})^2+B_2(k^{y'})^2+B_3k^{z'}+B_4k^{y'}+B_5k^{z'}k^{y'},
\end{equation}
where the $B$'s are functions of $x^{\mu'}$ and $|\mathbf{k}'|$ (since the $k$'s are just linear combinations of spherical harmonics, this method yields the same result as proceeding in the more rigorous fashion of expanding the functions in terms of spherical harmonics). Inserting this ansatz into $\delta{f_1}\big{|}_{coll}$, performing the integrals and comparing powers of $k$ on both sides, we find

\begin{equation}
B_1 = \frac{10}{9}\frac{1}{n'\sigma_T}\frac{1}{|\mathbf{k}'|}\frac{\partial{f_0}}{\partial|\mathbf{k}'|}\nabla_{z'}U_{z'} \label{B1},
\end{equation}
\begin{equation}
B_2 = \frac{10}{9}\frac{1}{n'\sigma_T}\frac{1}{|\mathbf{k}'|}\frac{\partial{f_0}}{\partial|\mathbf{k}'|}\nabla_{y'}U_{y'},
\end{equation}
\begin{equation}
B_3 = \frac{1}{n'\sigma_T}\bigg{(}\frac{\partial{f_0}}{\partial{|\mathbf{k'}|}}\nabla_{0'}U_{z'}-\frac{1}{|\mathbf{k}'|}\frac{\partial{f_0}}{\partial{z'}}\bigg{|}_{|\mathbf{k}'|}\bigg{)},
\end{equation}
\begin{equation}
B_4 = \frac{1}{n'\sigma_T}\bigg{(}\frac{\partial{f_0}}{\partial{|\mathbf{k'}|}}\nabla_{0'}U_{y'}-\frac{1}{|\mathbf{k}'|}\frac{\partial{f_0}}{\partial{y'}}\bigg{|}_{|\mathbf{k}'|}\bigg{)},
\end{equation}
\begin{equation}
B_5 = \frac{10}{9}\frac{1}{n'\sigma_T}\frac{1}{|\mathbf{k'}|}\frac{\partial{f_0}}{\partial{|\mathbf{k}|}}\bigg{(}\nabla_{y'}U_{z'}+\nabla_{z'}U_{y'}\bigg{)} \label{B5}.
\end{equation}
In addition, however, we find that there are isotropic terms, i.e., only dependent on $|\mathbf{k}'|$, that arise from the integrations over $(k^{z'})^2$ and $(k^{y'})^2$; it is also apparent that $k^{0'}\partial{f_0}/\partial{t'}$, the first term in the sum in equation \eqref{boltzmannlhs}, represents an isotropic contribution to the left-hand side. Since the collision integral is zero for any isotropic term, however, these additional terms cannot be accounted for with our correction to the distribution function. (Equivalently, there are terms proportional to the $Y^{0}_{0}$ spherical harmonic, a constant, that cannot be balanced by adding more terms to the distribution function.) We are therefore forced to equate these extraneous collision terms and the time derivative, yielding an extra constraint that the zeroth-order distribution function must satisfy:

\begin{equation}
\frac{1}{3}\nabla_{i'}U^{i'}|\mathbf{k}'|\frac{\partial{f_0}}{\partial|\mathbf{k}'|} = \frac{\partial{f_0}}{\partial{t'}}\bigg{|}_{|\mathbf{k}'|} \label{entropy}.
\end{equation}
What does this condition mean physically? As an illustrative example, let us take the case where the zeroth-order distribution function is given by that for blackbody radiation:

\begin{equation}
f_0 = \frac{C}{e^\frac{{|\mathbf{k}'|}}{T}-1},
\end{equation}
where $C$ is a constant, the value of which is unimportant, and we have taken Boltzmann's constant to be one. With this form for $f_0$, equation \eqref{entropy} becomes

\begin{equation}
\frac{\partial{T}}{\partial{t'}} = -\frac{1}{3}T\,\nabla_{i'}U^{i'}.
\end{equation}
This, however, is just the gas energy equation for an isentropic, relativistic gas in the frame comoving with the fluid. Equation \eqref{entropy}, therefore, is equivalent to the statement that the zeroth-order distribution function be isentropic. 

The first-order correction to the distribution function, using equations \eqref{B1} -- \eqref{B5}, is thus

\begin{multline}
\rho'\kappa{}f_1 = \frac{10}{9}\frac{1}{|\mathbf{k}'|}\frac{\partial{f_0}}{\partial{|\mathbf{k}'|}}\bigg{(}(k^{z'})^2\nabla_{z'}U_{z'}+(k^{y'})^2\nabla_{y'}U_{y'}+k^{y'}k^{z'}(\nabla_{y'}U_{z'}+\nabla_{z'}U_{y'})\bigg{)} \\ +\frac{\partial{f_0}}{\partial{|\mathbf{k}'|}}\bigg{(}k^{z'}\nabla_{0'}U_{z'}+k^{y'}\nabla_{0'}U_{y'}\bigg{)}-\frac{1}{|\mathbf{k}'|}\bigg{(}k^{z'}\frac{\partial{f_0}}{\partial{z'}}\bigg{|}_{|\mathbf{k}'|}+k^{y'}\frac{\partial{f_0}}{\partial{y'}}\bigg{|}_{|\mathbf{k}'|}\bigg{)}
\end{multline}
where $\kappa = n'\sigma_T/\rho'$ is the opacity. Since we used a relativistic approach to derive this expression, we should be able to write it in a covariant fashion. This is indeed the case, the covariant form being

\begin{equation}
\rho'\kappa{f_1}= \frac{10}{9}\frac{1}{|\mathbf{k}'|}\frac{\partial{f_0}}{\partial{|\mathbf{k}'|}}\Pi^{\mu'}_{\,\,\sigma'}\Pi^{\nu'}_{\,\,\alpha'}k^{\sigma'}k^{\alpha'}\nabla_{\mu'}U_{\nu'}+\frac{\partial{f_0}}{\partial{|\mathbf{k}'|}}\Pi^{\nu'}_{\,\,\sigma'}k^{\sigma'}U^{\mu'}\nabla_{\mu'}U_{\nu'}-\frac{1}{|\mathbf{k}'|}\Pi^{\mu'}_{\,\,\nu'}k^{\nu'}\nabla_{\mu',|\mathbf{k}'|}f_0\label{f1eq1},
\end{equation}
where $\Pi^{\mu'\nu'} = U^{\mu'}U^{\nu'}+g^{\mu'\nu'}$ is the projection tensor introduced by \citet{eck40} (see section 1). We introduced the quantity $\nabla_{\mu',|\mathbf{k}'|}$ to signify the derivative with respect to coordinate $x^{\mu'}$ holding $|\mathbf{k}'|$ constant. This is an important distinction if we want to calculate derivatives of $f_1$ as we must use the relativistically-correct definition $|\mathbf{k}'| = \sqrt{g_{i'j'}k^{i'}k^{j'}}$ and $g_{i'j'}$ depends on $x^{\mu'}$. 

The preceding analysis only considered a single fluid element. However, the location of our origin, tantamount to the position of the fluid element under consideration, is arbitrary, meaning that equation \eqref{f1eq1} is applicable to the entire fluid. Also, even though we only considered planar flows, we can easily generalize the approach to include three-dimensional motion and variations, and \eqref{f1eq1} still holds. Greek indices therefore range from $0$ to 3 in equation \eqref{f1eq1}.

In the next section we will use equation \eqref{f1eq1} and integrals thereof to write the equations of radiation hydrodynamics, equation \eqref{radhydro}, in terms of the four-velocity of the fluid, the mass density, and the radiation energy density, to which we will add the continuity equation for the scatterers to close the system. How do we reconcile these with equation \eqref{entropy}, which seems to be an additional constraint? Recall that equation \eqref{entropy} was derived by equating the ``extra," isotropic terms arising from both the collision integral and the derivatives of $f_0$. If we were to attempt to derive the \emph{second} order correction to the distribution function, $f_2$, we would doubtless encounter more isotropic terms arising from the derivatives of $f_1$ and collision integrals of $f_2$, this time to first order in the optical depth, which we would have to add to equation \eqref{entropy}. Equation \eqref{entropy} therefore only captures effects to zeroth order in the mean free path. Although it constrains the spatial and temporal derivatives of $f_0$ appearing in equation \eqref{f1eq1}, it has no effect on the form of this equation. However, the gas energy equation (see equation \eqref{gasenergyco}) should reduce to equation \eqref{entropy} in the limit that $\rho'\kappa \rightarrow \infty$.

\section{Relativistic, diffusive equations of radiation hydrodynamics}
The equations of radiation hydrodynamics, valid in any frame, are given by equation \eqref{radhydro}. Because we now have the distribution function to the requisite order in the optical depth, we can simplify those equations by writing the radiation energy-momentum tensor as $R^{\mu\nu} = R^{\mu\nu}_0+R^{\mu\nu}_1$, where each $R^{\mu\nu}$ is given by equation \eqref{radmom} with the appropriate distribution function. We will derive each of these tensors in the comoving frame but write them in a manifestly covariant form, the results then being applicable in any coordinate system.

The comoving, isotropic energy-momentum tensor is given by

\begin{equation*}
R_0^{\mu'\nu'} = \int{k^{\mu'}k^{\nu'}f_0\frac{d^3k'}{k^{0'}}},
\end{equation*}
which we can show is equivalent to

\begin{equation}
R_0^{\mu'\nu'} = e'U^{\mu'}U^{\nu'}+\frac{1}{3}e'\Pi^{\mu'\nu'} \label{radtens0},
\end{equation}
where

\begin{equation}
e' = 4\pi\int_0^{\infty}f_0|\mathbf{k}'|^3d|\mathbf{k}'| \label{e'}
\end{equation}
is the isotropic radiation energy density. Since the projection tensor is orthogonal to the four-velocity, a time-like vector that selects the energy component of a tensor, it is reasonable to associate $e'/3$ with the pressure, or momentum density, exhibited by the fluid. With this association, equation \eqref{radtens0}, not surprisingly, demonstrates that radiation acts like a gas with a relativistic equation of state, i.e., one with an adiabatic index of $4/3$. 

The correction to the energy-momentum tensor,

\begin{equation}
R_1^{\mu'\nu'} = \int{}k^{\mu'}k^{\nu'}f_1\frac{d^3k'}{k^{0'}},
\end{equation}
will have a number of terms, as evidenced by equation \eqref{f1eq1}. Integrating by parts and performing some simple manipulations, we can show that $R_1^{\mu'\nu'}$ is given by

\begin{multline}
\rho'\kappa{}R^{\mu'\nu'}_1 = -\frac{10}{9}\frac{4}{3}e'U^{\mu'}U^{\nu'}\nabla_{\alpha'}U^{\alpha'}-\frac{10}{9}\frac{4}{15}e'\Pi^{\mu'\sigma'}\Pi^{\nu'\rho'}\bigg{(}\nabla_{\sigma'}U_{\rho'}+\nabla_{\rho'}U_{\sigma'}-\frac{2}{3}g_{\sigma'\rho'}\nabla_{\alpha'}U^{\alpha'}\bigg{)} \\ -\frac{10}{9}\frac{4}{9}e'\Pi^{\mu'\nu'}\nabla_{\alpha'}U^{\alpha'}-\frac{1}{3}\bigg{(}U^{\mu'}\Pi^{\nu'\sigma'}+U^{\nu'}\Pi^{\mu'\sigma'}\bigg{)}\bigg{(}4e'U^{\rho'}\nabla_{\rho'}U_{\sigma'}+\nabla_{\sigma'}e'\bigg{)} \label{comovingradtens}.
\end{multline}
Written in this manner, it is evident that $R^{\mu\nu}_1$ transforms like a tensor. It should also be noted that each term in the tensor contains a derivative of a quantity, either the fluid velocity or the energy of the radiation field, with respect to the optical depth, which is what we expected. 

Equation \eqref{comovingradtens} differs from equation \eqref{deltaT}, the viscous stress tensor proposed by \citet{wei71}, in two notable ways. The first is that we have not postulated the existence of a temperature; instead we just used the fact that the zeroth-order distribution function is isotropic in the comoving frame to leave the comoving energy density, given by equation \eqref{e'}, as an unknown. The temperature has in fact been a difficult quantity to define in past treatments (see Weinberg's discussion of the reconciliation between the results of \citet{tho30} and Eckart's general form for a relativistic viscous stress tensor; see also \citealt{lim90}), and it is reassuring to find that the physics is perfectly well-described without invoking such a quantity. 

The second difference is contained in the presence of the first term of our stress tensor, proportional to $U^{\mu}U^{\nu}$, which shows that there is a correction to the comoving radiation energy density -- \emph{defined} to be zero by \citet{eck40} and \citet{wei71} -- given by

\begin{equation}
\Delta{e'} = -\frac{10}{9}\frac{4}{3}\frac{e'}{\rho'\kappa}\nabla_{\alpha'}U^{\alpha'} \label{deltae}.
\end{equation}
This expression can be understood as follows: imagine that we take a spherical volume of fluid and contract it by some amount, i.e., such that $\nabla_{\alpha'}U^{\alpha'}$ is negative. Because of the nature of the Thomson cross section, any radiation intersected by the contracting fluid will be scattered preferentially in the direction of motion. Therefore, the radiation energy in this volume will be increased owing to the in-scattering of photons, which is reflected in equation \eqref{deltae}. Furthermore, if we recall that the term in the stress tensor proportional to $\Pi^{\mu'\nu'}$ can be interpreted as the pressure exerted by the radiation on a fluid element, we see that

\begin{equation}
\Delta{p}' = -\frac{10}{9}\frac{4}{9}\frac{e'}{\rho'\kappa}\nabla_{\mu'}U^{\mu'} = \frac{1}{3}\Delta{e'}.
\end{equation}
This demonstrates that the change in pressure is 1/3 the change in energy, which is what we expect -- the relativistic nature of the photon gas is preserved independent of the manner in which we expand the distribution function. It can also be shown that the trace of $R^{\mu\nu}_1$ is zero, which is another familiar property of a relativistic gas.

Comparing the other terms in equation \eqref{comovingradtens} to the form of an arbitrary viscous stress tensor given by \eqref{deltaT}, we find

\begin{equation}
\eta = \frac{8}{27}\frac{e'}{\rho'\kappa} \label{etaeq}
\end{equation}
for the coefficient of dynamic viscosity, which agrees with the findings of \citet{loe92}. \citet{tho30} used an incorrect form for the Thomson cross section, so \citet{wei71} did not have the factor of 10/9. The proportionality to $e'$ is sensible, as the viscous effect is mediated by radiation; therefore, a higher radiation energy density permits a higher transfer of energy and momentum to neighboring fluid elements. The viscous effect is also proportional to the mean free path of the radiation, which is also a reasonable result: smaller mean free paths mean that the observed velocity difference across a mean free path is smaller for a given shear, implying less transfer of momentum per scattering. The coefficient of bulk viscosity is found to be

\begin{equation}
\zeta = \frac{1}{3}\eta,
\end{equation}
and is not zero, as predicted by \citet{eck40} and \citet{wei71} for a radiation-dominated fluid. Because we did not introduce a temperature, we cannot define a coefficient of heat conduction in a manner analogous to that of \citet{wei71}. However, for the case where the zeroth-order distribution function is that of blackbody radiation, it can be verified that $\chi = 4aT^3/3$, which agrees with his findings. 

We can also calculate the correction to the flux of photons, where the flux four-vector is given by

\begin{equation}
F^{\mu} = \int{k^{\mu}f\frac{d^3k}{k^{0}}}.
\end{equation}
As expected, the zeroth-order flux only has a non-vanishing number density in the comoving frame, given by

\begin{equation}
F_0^{0'} \equiv N' = 4\pi\int{}f_0|\mathbf{k'}|^2d|\mathbf{k}'|,
\end{equation}
which can be written covariantly as

\begin{equation}
F^{\mu'}_0 = N'U^{\mu'}.
\end{equation}
As for the energy-momentum tensor, we can use equation \eqref{f1eq1} and integrate by parts to write the correction to the flux vector in terms of $N'$. We find

\begin{equation}
\rho'\kappa{}F^{\mu'}_1 = -\frac{10}{9}N'U^{\mu'}\nabla_{\alpha'}U^{\alpha'}-\frac{1}{3}\Pi^{\mu'\sigma'}\bigg{(}3N'U^{\alpha'}\nabla_{\alpha'}U_{\sigma'}+\nabla_{\sigma'}N'\bigg{)} \label{fluxcorr}.
\end{equation}
We see that this expression yields

\begin{equation}
\Delta{N'} = -U_{\mu'}F^{\mu'}_1 = -\frac{10}{9}\frac{N'}{\rho'\kappa}\nabla_{\alpha'}U^{\alpha'} \label{deltaN}
\end{equation}
as the correction to the comoving number density of photons, a result in contrast to the analysis of \citet{eck40}, who \emph{defined} $\Delta{N'}$ to be zero. However, equation \eqref{deltaN} has a similar interpretation to equation \eqref{deltae}: by noting that a contracting gas preferentially scatters photons in the direction of motion of the scatterers, one would expect an increased amount of radiation within that contracting volume. We also find

\begin{equation}
\frac{\Delta{e'}}{e'} = \frac{4}{3}\frac{\Delta{N'}}{N'},
\end{equation}
reaffirming the notion that radiation behaves as a relativistic gas and demonstrating that it is these extra photons, $\Delta{N'}$, that add to the energy of the contracting fluid. 

For the case of a cold gas, where $T^{\mu\nu} = \rho'U^{\mu}U^{\nu}$, we will, for completeness, write down the full set of equations:

\begin{equation}
\nabla_{\mu}\bigg{(}\rho'U^{\mu}\bigg{)} = 0 \label{masscontco},
\end{equation}
\begin{multline}
\nabla_{\mu}\bigg{[}\bigg{\{}\rho'+\frac{4}{3}e'\bigg{(}1-\frac{10}{9}\frac{1}{\rho'\kappa}\nabla_{\alpha}U^{\alpha}\bigg{)}\bigg{\}}U^{\mu}U^{\nu}\bigg{]}-\frac{8}{27}\nabla_{\mu}\bigg{[}\frac{e'}{\rho'\kappa}\Pi^{\mu\sigma}\Pi^{\nu\beta}\bigg{(}\nabla_{\sigma}U_{\beta}+\nabla_{\beta}U_{\sigma}+g_{\beta\sigma}\nabla_{\alpha}U^{\alpha}\bigg{)}\bigg{]} \\ 
+\frac{1}{3}g^{\mu\nu}\partial_{\mu}e'-\frac{1}{3}\nabla_{\mu}\bigg{[}\frac{e'}{\rho'\kappa}\bigg{(}\Pi^{\mu\sigma}U^{\nu}+\Pi^{\nu\sigma}U^{\mu}\bigg{)}\bigg{(}4U^{\beta}\nabla_{\beta}U_{\sigma}+\partial_{\sigma}\ln{}e'\bigg{)}\bigg{]} = 0 \label{radhydroco}.
\end{multline}
The first of these is just the continuity equation. We will also derive the gas energy equation, obtained by contracting equation \eqref{radhydroco} with the four velocity, which gives

\begin{multline}
\nabla_{\mu}(e'U^{\mu})+\frac{1}{3}e'\nabla_{\mu}U^{\mu} = \frac{4}{3}\frac{10}{9}\nabla_{\mu}\bigg{[}\frac{e'}{\rho'\kappa}U^{\mu}\nabla_{\alpha}U^{\alpha}\bigg{]}+\frac{8}{27}\frac{e'}{\rho'\kappa}\bigg{(}\nabla_{\sigma}U_{\beta}+\nabla_{\beta}U_{\sigma}+g_{\sigma\beta}\nabla_{\alpha}U^{\alpha}\bigg{)}\Pi^{\mu\sigma}\nabla_{\mu}U^{\beta} \\
+\frac{1}{3}\Pi^{\mu\sigma}\nabla_{\mu}\bigg{[}\frac{e'}{\rho'\kappa}\bigg{(}4U^{\beta}\nabla_{\beta}U_{\sigma}+\partial_{\sigma}\ln{e'}\bigg{)}\bigg{]}+\frac{1}{3}\frac{e'}{\rho'\kappa}\bigg{(}4U^{\beta}\nabla_{\beta}U_{\sigma}+\partial_{\sigma}\ln{e'}\bigg{)}\bigg{(}2U^{\mu}\nabla_{\mu}U^{\sigma}+U^{\sigma}\nabla_{\mu}U^{\mu}\bigg{)}
\label{gasenergyco}.
\end{multline}
The left-hand side is just the change in energy for an adiabatic, $\gamma = 4/3$ gas, where $\gamma$ is the adiabatic index. Note that if $e' \sim T^4$ and we let $\rho'\kappa \rightarrow \infty$, the left-hand side equals equation \eqref{entropy}. The right-hand side therefore represents the energy added to the radiation during interactions with the scatterers (see section 6 for a discussion concerning the entropy generated by this heat addition). 

\section{Perturbation analysis}
\citet{his85} showed that the general viscous tensor proposed by \citet{eck40} is unstable to small perturbations in a fluid. An interesting question is whether or not these instabilities appear in our set of equations.

To answer this question, consider an equilibrium solution $(\rho',e',U^{\mu})$ where all of the variables are constants in space and time, the fluid is motionless and the space is flat. On top of this equilibrium solution we will impose perturbations on our variables $(\delta\rho',\delta{e'},\delta{U^{\mu}})$ small enough such that their products are negligible. With this configuration, the zeroth-order fluid equations are trivially satisfied. The first-order perturbations to the energy-momentum tensors are

\begin{equation}
\delta{T^{\mu\nu}} = U^{\mu}U^{\nu}\delta\rho'+\rho'(U^{\nu}\delta{U^{\mu}}+U^{\mu}\delta{U^{\nu}}),
\end{equation}
\begin{equation}
\delta{R^{\mu\nu}_0} = \frac{4}{3}U^{\mu}U^{\nu}\delta{e'}+\frac{4}{3}e'(U^{\nu}\delta{U}^{\mu}+U^{\mu}\delta{U}^{\nu})+\frac{1}{3}g^{\mu\nu}\delta{e'},
\end{equation}
\begin{multline}
\delta{R^{\mu\nu}_1} = -\frac{10}{9}\frac{4}{3}\frac{e'}{\rho'\kappa}U^{\mu}U^{\nu}\nabla_{\alpha}\delta{U^{\alpha}}-\frac{8}{27}\frac{e'}{\rho'\kappa}\Pi^{\mu\sigma}\Pi^{\nu\rho}\bigg{(}\nabla_{\sigma}\delta{U_{\rho}}+\nabla_{\rho}\delta{U_{\sigma}}-\frac{2}{3}g_{\sigma\rho}\nabla_{\alpha}\delta{U^{\alpha}}\bigg{)} \\ 
-\frac{10}{9}\frac{4}{9}\frac{e'}{\rho'\kappa}\Pi^{\mu\nu}\nabla_{\alpha}\delta{U^{\alpha}}-\frac{1}{3}\bigg{(}U^{\mu}\Pi^{\nu\sigma}+U^{\nu}\Pi^{\mu\sigma}\bigg{)}\bigg{(}4\frac{e'}{\rho'\kappa}U^{\rho}\nabla_{\rho}\delta{U}_{\sigma}+\frac{1}{\rho'\kappa}\nabla_{\sigma}\delta{e'}\bigg{)}.
\end{multline}
We assumed here that the perturbations are small enough such that gravitational corrections can be ignored, i.e., $\delta{g_{\mu\nu}} = 0$. The first-order conservation equations that must be satisfied are now

\begin{equation}
\nabla_{\mu}(\delta{T}^{\mu\nu} +\delta{R}^{\mu\nu}_0)= -\nabla_{\mu}\delta{R}^{\mu\nu}_1 \label{deltaradhydro}.
\end{equation}
These four equations must also be coupled to the mass continuity equation, the first-order correction to which is

\begin{equation}
\nabla_{\mu}(U^{\mu}\delta\rho'+\rho'\delta{U^{\mu}}) = 0.
\end{equation} 
The normalization of the four-velocity, $U_{\mu}U^{\mu} = -1$, demonstrates that $\delta{U^{0}} = 0$.

For the present analysis we will restrict our attention to planar flows, such that $\delta{U^{x}} \equiv 0$ and any perturbations in the $x$ direction are exactly zero. In this case, only the $\nu = 0, y, $ and $z$ components of equation \eqref{deltaradhydro} are non-trivial. Carrying out the derivatives, we find that they become, respectively,

\begin{equation}
\frac{\partial}{\partial{t}}\delta{e'}+\frac{4}{3}e'\nabla_{i}\delta{U}^{i} = \frac{4}{3}(1+\frac{10}{9})\frac{e'}{\rho'\kappa}\frac{\partial}{\partial{t}}\nabla_{i}\delta{U}^{i}+\frac{1}{3}\frac{1}{\rho'\kappa}\nabla_{i}\nabla^{i}\delta{e'} \label{deltaeeq},
\end{equation}
\begin{equation}
(\rho'+\frac{4}{3}e')\frac{\partial}{\partial{t}}\delta{U^{z}}+\frac{1}{3}\frac{\partial}{\partial{z}}\delta{e'} = \frac{8}{27}\frac{e'}{\rho'\kappa}\bigg{(}\nabla_{i}\nabla^{i}\delta{U}^{z}+2\nabla_{i}\nabla_{z}\delta{U}^{i}\bigg{)}+\frac{4}{3}\frac{e'}{\rho'\kappa}\frac{\partial^2}{\partial{t}^2}\delta{U}^{z}+\frac{1}{3}\frac{1}{\rho'\kappa}\frac{\partial}{\partial{t}}\frac{\partial}{\partial{z}}\delta{e'},
\end{equation}
\begin{equation}
(\rho'+\frac{4}{3}e')\frac{\partial}{\partial{t}}\delta{U^{y}}+\frac{1}{3}\frac{\partial}{\partial{y}}\delta{e'} = \frac{8}{27}\frac{e'}{\rho'\kappa}\bigg{(}\nabla_{i}\nabla^{i}\delta{U}^{y}+2\nabla_{i}\nabla_{y}\delta{U}^{i}\bigg{)}+\frac{4}{3}\frac{e'}{\rho'\kappa}\frac{\partial^2}{\partial{t}^2}\delta{U}^{y}+\frac{1}{3}\frac{1}{\rho'\kappa}\frac{\partial}{\partial{t}}\frac{\partial}{\partial{y}}\delta{e'} \label{deltamomyeq}.
\end{equation}

We will now impose the restriction that each of our perturbations varies periodically as

\begin{equation}
\delta\xi = \xi_0e^{ik_{\mu}x^{\mu}},
\end{equation}
where $k_{\mu} = (-\omega,k_{i})$ and $x^{\mu} = (t,x^{i})$ (this approach is equivalent to taking the Fourier transform of the equations). Before substituting these expressions into equations \eqref{deltaeeq} -- \eqref{deltamomyeq}, however, first recall that $\delta{}R^{\mu\nu}_1 < \delta{}R^{\mu\nu}_0$, as required by our ordering scheme adopted when solving the Boltzmann equation perturbatively. Investigating the correction to the comoving energy density, this inequality implies

\begin{equation*}
\frac{e'}{\rho'\kappa}\nabla_{i}\delta{U^{i}} \lesssim\delta{e'},
\end{equation*}
or, in terms of the wavenumber, 

\begin{equation*}
\bigg{|}\frac{k_i\delta{U^{i}}}{\rho'\kappa}\bigg{|} \lesssim \frac{\delta{e'}}{e'},
\end{equation*}
where we introduced the absolute value signs because of the presence of the imaginary unit. Since $\delta{e'} \sim e'$, $\delta{U}^{i} \sim 1$ and $k_i \sim 1/\lambda$, where $\lambda$ is the wavelength of the perturbation, this inequality becomes $\lambda\rho'\kappa \gtrsim 1$. We see, therefore, that the optical depth across one wavelength must be greater than one in order for this perturbation analysis to hold. Equivalently, the wavelength of the perturbation, or the scale over which the perturbation acts, must be larger than the mean free path of a photon. If we induce changes on spatial scales smaller than the mean free path, we will violate the assumption that the mean free path is a good ``smallness" parameter for describing interactions between the fluid and the radiation. 

Investigating the other terms in $\delta{}R^{\mu\nu}_1$, we find that the inequality

\begin{equation*}
\frac{e'}{\rho'\kappa}U^{\rho}\nabla_{\rho}\delta{U}^{\mu} \lesssim U^{\mu}\delta{e'}
\end{equation*}
must also be upheld, which can be rearranged to give

\begin{equation}
\frac{\omega}{\rho'\kappa} \lesssim 1.
\end{equation}
Recalling that $1/\omega$ is proportional to the period of oscillation of the perturbation, this inequality means that the oscillation time scale must be longer than the light-crossing time over the mean free path. 

With the previous inequalities in mind, we will now insert our Fourier modes into equations \eqref{deltaeeq} -- \eqref{deltamomyeq}. The resulting three algebraic relations can be written as

\begin{equation}
M^{\mu}_{\,\,\nu}\delta\xi^{\nu} = 0,
\end{equation}
where

\begin{equation}
\delta\xi^{\nu} = \left(
\begin{array}{c}
\delta{e'} \\
\delta{U^{z}} \\
\delta{U^{y}}
\end{array} \right)
\end{equation}
and

\begin{equation}
M^{\mu}_{\,\,\nu}
= \left(
\begin{array}{ccc}
-i\omega+\frac{1}{3}\frac{1}{\rho'\kappa}k^2 & \frac{4}{3}\bigg{(}ie'-\frac{19}{9}\frac{e'}{\rho'\kappa}\omega\bigg{)}k^{z} & \frac{4}{3}\bigg{(}ie'-\frac{19}{9}\frac{e'}{\rho'\kappa}\omega\bigg{)}k^{y} \\
\frac{1}{3}\bigg{(}i-\frac{\omega}{\rho'\kappa}\bigg{)}k^{z} & -i\omega(\rho'+\frac{4}{3}e')+\frac{8}{27}\frac{e'}{\rho'\kappa}\bigg{(}\tilde{k}^2+2(k^{z})^2\bigg{)} & \frac{16}{27}\frac{e'}{\rho'\kappa}k^{z}k^{y} \\
\frac{1}{3}\bigg{(}i-\frac{\omega}{\rho'\kappa}\bigg{)}k^{y} & \frac{16}{27}\frac{e'}{\rho'\kappa}k^{z}k^{y} & -i\omega(\rho'+\frac{4}{3}e')+\frac{8}{27}\frac{e'}{\rho'\kappa}\bigg{(}\tilde{k}^2+2(k^{y})^2\bigg{)}
\end{array}\right), \label{Mmatrix}
\end{equation}
where $k^2 \equiv (k^{y})^2+(k^{z})^2$ and $\tilde{k}^2 \equiv k^2+9\omega^2/2$.

If the perturbations $\delta\xi^{\nu}$ are to be non-trivial, we demand det$(M^{\mu}_{\,\,\nu}) = 0$, which results in a dispersion relation that gives $w(k^i)$. Before setting the entire determinant equal to zero, recall that physically-meaningful frequencies and wavenumbers (from the standpoint that they can be described as viscous corrections to the energy momentum tensor) satisfy $w/(\rho'\kappa) \sim k^{i}/(\rho'\kappa) \lesssim 1$. Therefore, a first approximation to the dispersion relation can be obtained by setting all appearances of $1/(\rho'\kappa)$ to zero in the matrix given by \eqref{Mmatrix}. Doing so and taking the determinant, we find

\begin{equation*}
\det(M^{\mu}_{\,\,\nu}) = i\omega\bigg{(}\rho'+\frac{4}{3}e'\bigg)\bigg(\omega^2\bigg(\rho'+\frac{4}{3}e'\bigg)-\frac{4}{9}e'k^2\bigg) = 0,
\end{equation*}
the solutions to which are clearly

\begin{equation}
\omega = 0, \quad \omega = \pm\frac{2}{3}\sqrt{\frac{e'}{\rho'+\frac{4}{3}e'}}k \label{omega0}.
\end{equation}
The first of these represents a perturbation with a constant offset from the surrounding medium. The second two are the sound waves that propagate through the optically-thick plasma; if $e' \ll \rho'$, we recover the familiar result for the sound speed of a radiation-dominated gas $c_s = \sqrt{4p'/3\rho'}$, where $p' = 3e'$ is the radiation pressure, while for $e' \gg \rho'$, we find $c_s = c/\sqrt{3}$, which is the correct ultrarelativistic speed of propagation. 

We can show that the full determinant can be written 

\begin{multline}
\det(M^{\mu}_{\,\,\nu}) = \bigg{\{}{\frac{8}{27}\frac{e'}{\rho'\kappa}\tilde{k}^2}-i\omega\bigg{(}\rho'+\frac{4}{3}e'\bigg{)}\bigg{\}} \\  \times \bigg{\{}\bigg{(}\frac{1}{3}\frac{k^2}{\rho'\kappa}-i\omega\bigg{)}\bigg{(}\frac{8}{27}\frac{e'}{\rho'\kappa}\bigg{(}\tilde{k}^2+2k^2\bigg{)}-i\omega\bigg{(}\rho'+\frac{4}{3}e'\bigg{)}\bigg{)}-\frac{4}{9}e'\bigg{(}i-\frac{19}{9}\frac{\omega}{\rho'\kappa}\bigg{)}\bigg{(}i-\frac{\omega}{\rho'\kappa}\bigg{)}k^2\bigg{\}},
\end{multline}
meaning that the dispersion relation is given by the solutions to

\begin{equation}
\frac{8}{27}\frac{e'}{\rho'\kappa}\bigg{(}k^2+\frac{9}{2}\omega^2\bigg{)}-i\omega\bigg{(}\rho'+\frac{4}{3}e'\bigg{)} = 0 \label{disp0},
\end{equation} 
\begin{equation}
\bigg{(}\frac{1}{3}\frac{k^2}{\rho'\kappa}-i\omega\bigg{)}\bigg{(}\frac{8}{9}\frac{e'}{\rho'\kappa}\bigg{(}k^2+\frac{3}{2}\omega^2\bigg{)}-i\omega\bigg{(}\rho'+\frac{4}{3}e'\bigg{)}\bigg{)}-\frac{4}{9}e'\bigg{(}i-\frac{19}{9}\frac{\omega}{\rho'\kappa}\bigg{)}\bigg{(}i-\frac{\omega}{\rho'\kappa}\bigg{)}k^2 = 0 \label{disp1},
\end{equation}
where in each of these we used the definition of $\tilde{k}^2$. 

To solve these, recall that the perturbation frequencies permissible in our analysis are approximately given by equation \eqref{omega0} with small corrections of order $1/(\rho'\kappa)$. If we set $\rho'\kappa = 0$ in the preceding two polynomials, we see that the first reduces to $\omega = 0$, while the second gives the sound waves. In \eqref{disp0}, we will therefore let $\omega = \omega_1/(\rho'\kappa)$ and keep only first order terms in $1/(\rho'\kappa)$. Doing so, we find

\begin{equation}
\omega_1 \simeq -i\frac{8}{27}\frac{e'}{\rho'+\frac{4}{3}e'}k^2 \label{omega11}.
\end{equation}
Since the perturbations scale as $\delta\xi^{\nu} \sim e^{-i\omega{t}}$, $\omega_1$ is a decaying solution. The e-folding timescale of constant-offset perturbations is thus

\begin{equation}
\tau_{d} \simeq \frac{27}{8}\frac{\rho'c^2+\frac{4}{3}e'}{e'c}\frac{\rho'\kappa}{k^2} \label{taudecay},
\end{equation}
where we explicitly reintroduced the speed of light $c$, which shows that smaller-scale fluctuations decay more rapidly. 

To solve equation \eqref{disp1}, we will let $\omega = \omega_{\pm}+\omega_1/(\rho'\kappa)$, where $\omega_{\pm}$ is given by

\begin{equation}
\omega_{\pm} =  \pm\frac{2}{3}\sqrt{\frac{e'}{\rho'+\frac{4}{3}e'}}k,
\end{equation}
the zeroth-order (in the mean free path) solution to the dispersion relation. Using this approximation and keeping only highest-order terms, we find

\begin{equation}
\omega_1 = -\frac{i}{6}\frac{(\rho')^2+\frac{32}{27}\rho'e'+\frac{32}{27}(e')^2}{(\rho'+\frac{4}{3}e')^2}k^2.
\end{equation}
The decay timescale for traveling waves is therefore on the same order as that for the constant-offset perturbations. 

When we solved these dispersion relations, we used the fact that $\omega/(\rho'\kappa)$ must be small for our analysis to hold. If we had not taken this perturbative approach, however, we would have obtained conflicting results. For example, the exact solution to equation \eqref{disp0} is

\begin{equation}
\omega = i\frac{\rho'+\frac{4}{3}e'}{\frac{8}{3}\frac{e'}{\rho'\kappa}}\bigg{(}1\pm\sqrt{1+\frac{128}{81}\bigg{(}\frac{\frac{e'}{\rho'\kappa}k}{\rho'+\frac{4}{3}e'}\bigg{)}^2}\bigg{)} \label{omegaex}.
\end{equation}
If we take the root with the negative sign, the result reduces to equation \eqref{omega11} in the limit that we keep only first-order corrections in $k/(\rho'\kappa)$; but, if we take the positive sign, we find that the solution is a \emph{growing} mode, showing that the fluid perturbations are unstable. However, since these growing modes always have $\omega/(\rho'\kappa) > 1$, our treatment of the equations of radiation hydrodynamics, which only considers fluid perturbations on timescales greater than the light-crossing time over one optical depth, is invalid. 

One can understand the physical origin of these growing modes by imagining that we oscillate a fluid parcel in such a way that its period is one-half of the light-crossing time over the mean free path that separates it from its neighboring fluid element. In this case, by the time the information from the neighboring fluid parcel returns back to the originally-perturbed parcel, the relative velocity between the two fluid elements will be in the same direction as the initial perturbation. The viscous force will then serve to increase the amplitude of the oscillation of the fluid element and, taken over many optical depth light-crossing times, this effect will only be amplified, resulting in a runaway process. 

It is for this reason that the analysis of \citet{his85} resulted in the prediction of growing modes. If we take the positive root of equation \eqref{omegaex} and use numbers for water at room temperature, we find $\tau = 1/\omega \simeq 10^{-35}\text{ s}$ is the e-folding time of the perturbations. As pointed out by \citet{his85}, viscous heating would cause water to boil on an absurdly short timescale. This result is incorrect, however, because of the inability of the fluid to communicate over such timescales; these high frequencies blatantly violate the assumption that the mean free path of the radiation is sufficiently small to describe local fluid deformations.

For a general coefficient of dynamic viscosity $\eta$, this requirement can be translated to a statement similar to, ``frequencies must satisfy the inequality $\omega\eta/e' \lesssim 1$, where $e'$ is the locally-observed energy density of the fluid, in order for viscous effects to describe physically the perturbations that take place in the medium." It is also not surprising that in the limit of $c \rightarrow \infty$, this result disappears as the light-crossing time is zero, meaning that the Newtonian limit of these equations will never suffer from such growing instabilities. We can see this explicitly by reinserting the factors of $c$ into equation \eqref{disp0}, which reads

\begin{equation*}
\frac{8}{27}\frac{e'}{\rho'\kappa}\bigg{(}k^2+\frac{9}{2}\frac{\omega^2}{c^2}\bigg{)}-\frac{i\omega}{c}\bigg{(}\rho'c^2+\frac{4}{3}e'\bigg{)} = 0.
\end{equation*}
Ignoring the factor of $\omega^2/c^2$ and solving reveals only decaying solutions.

\section{Discussion and Conclusions}
We have shown, in agreement with intuition and past efforts, that radiation behaves like a viscosity in the optically-thick limit. Our analysis is based on the general relativistic Boltzmann equation, general relativistic corrections being necessary because the scattering, for which we used the differential Thomson cross section, is handled most easily in the local, accelerating frame of the fluid. The correction to the photon distribution function, given explicitly by equation \eqref{f1eq1}, was used to calculate the correction to the energy-momentum tensor of the radiation (equation \eqref{comovingradtens}), and many of the terms agreed with the predictions of \citet{eck40} and \citet{wei71}.

There are, however, a few differences between our form for the correction to the radiation energy-momentum tensor and the general viscous stress tensor proposed by \citet{wei71}, the first being that ours does not depend on thermodynamic considerations such as the assignment of a temperature to the zeroth-order distribution function. The temperature-independence of our equations means that the isotropic radiation distribution need not correspond to that of blackbody radiation. Indeed, because our approach treats the radiation and the scatterers as two interacting media, it is not clear what a single temperature would mean. The second difference is that our viscous tensor contains a correction to the comoving energy density of the radiation, which is not predicted by previous approaches to deriving the viscous tensor for a relativistic fluid. We demonstrated that this term appears naturally from the manner in which radiation is scattered in a contracting (or expanding) fluid element. This correction is consistent with, and in fact expected from, the notion that the radiation is a relativistic (adiabatic index of 4/3) gas. Furthermore, our model predicts a correction to the number density of photons (equations \eqref{fluxcorr} and \eqref{deltaN}), the presence of which corresponds exactly with the increase in the energy density of the fluid. 

We performed a perturbation analysis on our equations, took the Fourier transform of the perturbed equations and  calculated the dispersion relation (equations \eqref{disp0} and \eqref{disp1}). Because our treatment is concerned with the limit in which the stress tensor is a small perturbation proportional to the mean free path of the radiation, the wavenumbers ($k^{i}$) and frequencies ($\omega$) of the Fourier modes must satisfy $k^{i}/(\rho'\kappa) \sim \omega/(\rho'\kappa) \lesssim 1$. These inequalities mean, sensibly, that fluid motions cannot alter the radiation field over scales smaller than the mean free path and at rates faster than the light-crossing time over the mean free path. With these inequalities in mind, we showed that the viscous terms cause the perturbations to decay exponentially, a familiar result. Interestingly, if one solves the dispersion relations without regard to the inequality $\omega/(\rho'\kappa) \lesssim 1$, one recovers growing mode solutions. This result sheds light on the results of \citet{his85}, who demonstrated that the \citet{eck40} relativistic viscous stress tensor is unstable to small perturbations. Our approach shows that such instabilities only arise when one violates the assumption that the viscous nature of the fluid is a ``good" approximation, i.e., when the light crossing time over the viscous length scale, defined in any problem by $\eta/(c\,e')$, $\eta$ being the coefficient of dynamic viscosity and $e'$ the locally-observed energy density of the fluid, is short enough to describe local fluid deformations. The unstable nature of the modes also vanishes in the non-relativistic limit as the light-crossing time over any distance is formally zero. 

\citet{wei71} calculated the bulk viscosity of a relativistic radiating fluid to be zero. Our analysis, on the contrary, determined the coefficient to be $\zeta = \eta/3$, which can be quite large for radiation-dominated plasmas. It might therefore be expected that our model would predict a different rate of entropy production in the universe. However, if we use his expression for the entropy production rate (see the expression above equation (2.20) in \citealt{wei71}) and confine our analysis to the case in which zeroth-order distribution function can be described by that of a blackbody, we find that our bulk viscosity and the correction to the comoving energy density exactly cancel. We therefore find that Weinberg's results concerning the entropy generation in the early universe by small anisotropies in the radiation field are upheld.

The previously-developed equations are applicable to a range of physical scenarios, the cosmological evolution of the early universe being one such application (see, in addition to \citealt{wei71}, \citealt{cad77, har77, tau78, joh88, hu02, kha12}). Another application would be in analyzing the physics of the boundary layers established between radiation-dominated jets and their environments \citep{ara92}, as viscous dissipation due to radiation could be important -- especially in the case where the motion of the outflow is relativistic \citep{wal90}. Such an analysis could shed light on the peculiar event \emph{Swift} J1644+57, thought to be the first-observed, jetted, super-Eddington tidal disruption event (TDE) \citep{bur11, blo11, can11, zau11}. If the transient jet carried away the accretion luminosity generated during the gravitational infall of tidally-stripped debris \citep{cou14}, the interaction between the jet and the overlying envelope may be well-described by the equations radiation hydrodynamics in the viscous limit. Comparing the theoretical results with observations could yield insight into the terminal Lorentz factor of the outflow, the properties of the tidally-disrupted star, and the mass of the black hole residing in the center of the host galaxy.

Applying these equations to the outflows predicted by the collapsar model of gamma-ray bursts (GRBs) \citep{woo93, mac99}, wherein the core of a massive, dying star collapses directly to a black hole from which a jet is launched, may also prove fruitful. If the mechanism that accelerates the outflow is ultimately derived from the prompt accretion of material onto the newly-formed black hole, i.e., if the fireball model correctly describes the dynamics \citep{ree92}, radiation would contribute significantly to the energetics of the jet. Even if the energy is provided by the spin of the black hole \citep{bla77} or the accretion disk \citep{bla82}, radiation could still play a large part in determining the dynamics of the outflow. An accurate representation of the interaction of the jet with the overlying stellar envelope and the circumstellar environment may therefore be obtained by employing the viscous equations of radiation hydrodynamics. A comparison of the theoretical expectation gleaned from such an analysis with the promptly-emitted gamma rays and the X-ray afterglow could yield new information concerning the progenitors of GRBs and their surrounding environments.

Owing to their apparent complexity, the equations of radiation hydrodynamics, when applied to supercritically accreting compact objects and their surroundings, are often solved numerically \citep{egg88, oku02, oku05, ohs05, ohs07}. Simulations of such systems have also been extended to include magnetic fields (radiation magnetohydrodynamic; RMHD), the presence of which is potentially important not only for the dynamics of the gas but also for MRI-induced accretion and jet collimation \citep{tur03, ohs09, sad14, mck14}. In these simulations, strong gravity is incorporated either through the pseudo-Newtonian potential of \citet{pac80}, or by using the covariant set of equations and the Kerr geometry. In all of these numerical schemes, the moment formalism is adopted (see the introduction) when solving for the properties of the radiation and incorporating the coupling of the radiation to the gas. In addition, a ``closure'' relation is adopted that allows one to truncate the number of moments needed to close the system of equations at a finite level. Given our preceding analysis, an interesting question is whether or not these closure relations accurately capture relativistic radiation viscosity.

Flux-limited diffusion \citep{lev81}, wherein the flux is proportional to the gradient of the energy density, is one such closure scheme used by, e.g., \citet{ohs05} and \citet{ohs09}. This approximation allows one to interpolate between optically-thick and thin regimes, useful for an optically-thick disk that launches an optically-thin jet -- the result observed in many of these simulations. An issue with this type of closure, especially when considering the present paper, is that the viscosity must be included separately and is not a direct result of the equations of radiation hydrodynamics. \citet{ohs05}, for example, used a modified-alpha prescription for the coefficient of viscosity, setting the coefficient of dynamic viscosity to $\eta \sim p/\Omega_K$, where $p$ is the total (gas plus radiation) pressure and $\Omega_K$ is the Keplerian velocity. This form for the viscosity clearly does not reduce to equation \eqref{etaeq} in the radiation-dominated, viscous limit. The modified-alpha prescription also does not incorporate relativistic effects, those due to strong gravity or high velocities, both of which could be important in the vicinity of the black hole and in regions of high shear between the jet and the inflated disk. It is therefore unlikely that the flux-limited diffusion closure scheme accurately reproduces the effects of radiation viscosity in the high-optical depth limit.

The M1 closure scheme \citep{lev84} assumes that there exists some reference frame, the radiation rest frame (not necessarily the same as fluid rest frame), in which the radiation flux vanishes and the pressure is one third of the energy density (the Eddington approximation), and can also interpolate between optically-thick and -thin regimes. This scheme has been extended to incorporate general relativistic effects \citep{sad13}, and has been employed by \citet{sad14} and \citet{mck14} to study the role of radiation on hyperaccreting black hole-disk systems. Because the frame in which the Eddington approximation applies is not the fluid frame, there are small corrections to the energy-momentum tensor of the radiation in the fluid rest-frame when the optical depth is large, which is what one expects. If one keeps the lowest-order corrections to the fluid-frame pressure tensor (see equation (34) of \citet{sad13}), the result is $P^{i'j'} \simeq \delta^{i'j'}e'/3+\mathcal{O}(F^2)$, where $F$ is the fluid-frame energy flux, which demonstrates that, indeed, the fluid-frame radiation pressure reduces to the Eddington approximation when the optical depth is large. In the viscous limit, the rest-frame energy flux is proportional to gradients of the energy density and the velocity over the mean free path of the radiation (see the $(0,i)$ components of our equation \eqref{comovingradtens} or note that the diffusion limit is $\mathbf{F} \sim \nabla{e'}$). The M1 scheme therefore indicates that the first-order deviation of the pressure, or the momentum flux, from the Eddington closure is proportional to the square of the derivative of fluid quantities over the optical depth. This result, however, is in direct contrast with the $(i,j)$ components of our equation \eqref{comovingradtens} that demonstrate, consistent with the manner in which viscosity usually operates, that the lowest-order correction to the momentum flux in the viscous limit is proportional to the gradient of fluid quantities over the optical depth, i.e., it is of the same order as the energy flux. The M1 closure therefore does not accurately reproduce the viscous transport of momentum in the high-optical depth limit. Due to the covariant manner in which the M1 scheme is incorporated and the self-consistent inclusion of the radiation terms (i.e., no ad-hoc form for the viscosity), higher-order effects of the anisotropic radiation field are likely well-reproduced in the interaction between the relativistic jetted outflows and the inflated torus observed in the simulations of \citet{sad14} and \citet{mck14}.

Finally, \citet{jia14} recently performed a non-relativistic, 3D, RMHD simulation of the gas around a $\sim 6M_{\astrosun}$ black hole. Contrary to the previously-mentioned authors, they did not assume a closure relation to compute the moments of the radiation energy-momentum tensor and instead determined the pressure of the radiation directly, offering an unbiased depiction of the manner in which radiation transfers energy and momentum between neighboring gas parcels. The non-relativistic nature of the simulation, however, may underestimate the viscous effects in the transition between the mildly-relativistic outflow observed in their simulation and the disk material.

Having written down the equations of radiation hydrodynamics in the viscous limit, the authors plan to pursue the applications of those equations to relativistic, radiative shear layers in a companion paper. In the process, we plan to quantitatively assess how the differences between our radiation-viscous fluid equations and those of \citet{wei71} affect the physical evolution of astronomical systems. 

\acknowledgements
This work was supported in part by NASA Astrophysics Theory Program grant NNX14AB37G, NSF grant AST-1411879, and NASA's Fermi Guest Investigator Program.  We thank Charles Gammie for drawing our attention to the work of \citealt{his85}.

\clearpage


\begin{thebibliography}{}
\bibitem[Arav \& Begelman(1992)]{ara92} Arav N., Begelman M.C., 1992, ApJ, 401, 125
\bibitem[Blandford \& Payne(1982)]{bla82} Blandford R.D., Payne D.G., 1982, MNRAS, 199, 883
\bibitem[Blandford \& Znajek(1977)]{bla77} Blandford R.D., Znajek R.L., 1977, MNRAS, 179, 433
\bibitem[Bloom et al.(2011)]{blo11} Bloom J.S., Giannios D., Metzger B.D., et al., 2011, Sci, 333, 203
\bibitem[Buchler(1979)]{buc79} Buchler J.R., 1979, JQSRT, 22, 293
\bibitem[Burrows et al.(2011)]{bur11} Burrows D.N., Kennea J.A., Ghisellini G., et al., 2011, Natur, 476, 421
\bibitem[Caderni \& Fabbri(1977)]{cad77} Caderni N., Fabbri R., 1977, PhLB, 69, 508
\bibitem[Cannizzo et al.(2011)]{can11} Cannizzo J.K., Troja E., Lodato G., 2011, ApJ, 742, 732
\bibitem[Castor(1972)]{cas72} Castor J.I., 1972, ApJ, 178, 779
\bibitem[Chen \& Spiegel(2000)]{che00} Chen X., Spiegel E.A., 2000, ApJ, 540, 1069
\bibitem[Coughlin \& Begelman(2014)]{cou14} Coughlin E.R., Begelman M.C., 2014, ApJ, 781, 82
\bibitem[Debbasch \& van Leeuwen(2009a)]{deb09a} Debbasch F., van Leeuwen W.A., 2009a, PhyA, 388, 1079
\bibitem[Debbasch \& van Leeuwen(2009b)]{deb09b} Debbasch F., van Leeuwen W.A., 2009b, PhyA, 388, 1818
\bibitem[Eckart(1940)]{eck40} Eckart C., 1940, PhRv, 58, 919
\bibitem[Eggum et al.(1988)]{egg88} Eggum G.E., Coroniti F.V., Katz J.I., 1988, ApJ, 330, 142
\bibitem[Harrison(1977)]{har77} Harrison E.R., 1977, VA, 20, 341
\bibitem[Hiscock \& Lindblom(1985)]{his85} Hiscock W.A., Lindblom L., 1985, PhRvD, 31, 725
\bibitem[Hsieh \& Spiegel(1976)]{hsi76} Hsieh S.H., Spiegel E.A., 1976, ApJ, 207, 244
\bibitem[Hu \& Dodelson(2002)]{hu02} Hu W., Dodelson S., 2002, ARA\&A, 40, 171
\bibitem[Jiang et al.(2014)]{jia14} Jiang Y., Stone J.M., Davis S.W., 2014, ApJ, 796, 106
\bibitem[Johri \& Sudharsan(1988)]{joh88} Johri V.B., Sudharsan R., 1988, PhLA, 132, 316
\bibitem[Khatri et al.(2012)]{kha12} Khatri R., Sunyaev R.A., Chluba J., 2012, A\&A, 543, 136
\bibitem[Landau \& Lifshitz(1958)]{lan58} Landau L., Lifshitz E.M., 1958, Fluid Mechanics (Addison-Wesley, Reading, Mass.)
\bibitem[Lima \& Waga(1990)]{lim90} Lima J.A.S., Waga I., 1990, PhLA, 144, 432
\bibitem[Lindquist(1966)]{lin66} Lindquist R.W., 1966, AnPhy, 37, 487
\bibitem[Levermore(1984)]{lev84} Levermore C.D., 1984, JQSRT, 31, 149
\bibitem[Levermore \& Pomraning(1981)]{lev81} Levermore C.D., Pomraning G.C., 1981, ApJ, 248, 321
\bibitem[Loeb \& Laor(1992)]{loe92} Loeb A., Laor A., 1992, ApJ, 384, 115
\bibitem[MacFadyen \& Woosley(1999)]{mac99} MacFadyen A.I., Woosley S.E., 1999, ApJ, 524, 262
\bibitem[McKinney et al.(2014)]{mck14} McKinney J.C., Tchekhovskoy A., S\,adowski A., et al., 2014, MNRAS, 441, 3177
\bibitem[Mihalas \& Mihalas(1984)]{mih84} Mihalas D., Mihalas B.W., 1984, Foundations of Radiation Hydrodynamics \\ (New York: Oxford Univ. Press)
\bibitem[Misner \& Sharp(1965)]{mis65} Misner C.W., Sharp D.H., 1965, PhL, 15, 279
\bibitem[Munier(1986)]{mun86} Munier A., 1986, PhRvD, 33, 2111
\bibitem[Ohsuga(2007)]{ohs07} Ohsuga K., 2007, PASJ, 59, 1033
\bibitem[Ohsuga et al.(2009)]{ohs09} Ohsuga K., Mineshige S., Mori M., et al., 2009, PASJ, 61L, 7
\bibitem[Ohsuga et al.(2005)]{ohs05} Ohsuga K., Mori M., Nakamoto T., et al., 2005, ApJ, 628, 368
\bibitem[Okuda(2002)]{oku02} Okuda T., 2002, PASJ, 54, 253
\bibitem[Okuda et al.(2005)]{oku05} Okuda T., Teresi V., Toscano E., et al., 2005, MNRAS, 357, 295
\bibitem[Paczy\'nski \& Wiita(1980)]{pac80} Paczy\'nski B., Wiita P.J., 1980, A\&A, 88, 23
\bibitem[Rees \& M\'esz\'aros(1992)]{ree92} Rees M.J., M\'esz\'aros P., 1992, MNRAS, 258, 41
\bibitem[S\k{a}dowski et al.(2013)]{sad13} S\k{a}dowski A., Narayan R., Tchekhovskoy A., et al., 2013, MNRAS, 429, 3533
\bibitem[S\k{a}dowski et al.(2014)]{sad14} S\k{a}dowski A., Narayan R., McKinney J.C., et al., 2014, MNRAS, 439, 503
\bibitem[Tauber(1978)]{tau78} Tauber G.E., 1978, Ap\&SS, 57, 163
\bibitem[Thomas(1930)]{tho30} Thomas L.H., 1930, QJMat, 1, 239
\bibitem[Turner et al.(2003)]{tur03} Turner N.J., Stone J.M., Krolik J.H., et al., 2003, ApJ, 593, 992
\bibitem[Walker(1990)]{wal90} Walker M.A., 1990, ApJ, 348, 668
\bibitem[Weinberg(1971)]{wei71} Weinberg S., 1971, ApJ, 168, 175
\bibitem[Woosley(1993)]{woo93} Woosley S.E., 1993, ApJ, 405, 273
\bibitem[Zauderer et al.(2011)]{zau11} Zauderer B.A., Berger E., Soderberg A.M., et al., 2011, Natur, 476, 425
\end{thebibliography}
\end{document}